\documentclass[conference,compsoc]{IEEEtran}
\usepackage{times}
\usepackage{soul}
\usepackage{url}
\usepackage[hidelinks]{hyperref}
\usepackage[utf8]{inputenc}
\usepackage[small]{caption}
\usepackage{graphicx}
\usepackage{amsmath}
\usepackage{amsthm}
\usepackage{booktabs}
\usepackage{algorithm}
\usepackage{algorithmic}
\usepackage[switch]{lineno}
\usepackage{multirow} 
\usepackage{listings}
\usepackage{amsfonts}
\usepackage{xcolor}
\usepackage{tcolorbox} 
\usepackage{listings}
\usepackage{multicol}
\usepackage{svg}
\usepackage{float}
\usepackage{subfig}
\usepackage{adjustbox}
\usepackage{booktabs}
\usepackage{tabularx}
\ifCLASSOPTIONcompsoc
  \usepackage[nocompress]{cite}
\else
  \usepackage{cite}
\fi
\ifCLASSINFOpdf
\else
\fi
\begin{document}
%

\title{AutoDecompiler: Reinforcement-Learning-Trained Binary Decompilation LLM \\ for Feedback-Driven Multi-Turn Refinement}


%

\author{
\IEEEauthorblockN{
{Peipei Liu}\IEEEauthorrefmark{1},
{Jian Sun}\IEEEauthorrefmark{1}, 
{Mingzhe Xing}\IEEEauthorrefmark{1}, 
{Yicheng Zeng}\IEEEauthorrefmark{1}, 
{Zhaoteng Yan}\IEEEauthorrefmark{1},
{Lixiao Zhang}\IEEEauthorrefmark{1}, 
{Li Chen}\IEEEauthorrefmark{1}, 
{Dan Li}\IEEEauthorrefmark{1}\IEEEauthorrefmark{2}
}
\IEEEauthorblockA{\IEEEauthorrefmark{1}Zhongguancun Laboratory, Beijing, China; \IEEEauthorrefmark{2}Tsinghua University, Beijing, China}
}

\maketitle

\newtcolorbox{graybox}{
    colback=gray!10,
    coltext=black,
    colframe=gray!30,
    boxrule=0.5pt,
    arc=4pt,
    boxsep=5pt,
    left=6pt,
    right=6pt,
    top=6pt,
    bottom=6pt
}

\begin{abstract}
Binary decompilation is fundamental to security tasks such as vulnerability discovery, malware inspection, and executable-only program understanding. Recent LLM-based decompilation methods have shown promising results, but most of them still follow a single-turn generation paradigm: given assembly code or decompiler-produced pseudo-code, the model generates one decompilation output and stops. As a result, the generated code may look readable and even compile successfully, while still deviating from the behavior of the original binary and misleading downstream security analysis.

This paper presents AutoDecompiler, a decompilation-specialized LLM trained with reinforcement learning (RL) to perform feedback-driven multi-turn binary decompilation. Instead of treating decompilation as one-shot code generation, AutoDecompiler formulates it as an iterative refinement process in which the LLM revises generated decompiled code based on compilation, execution, and input/output testing feedback. To train AutoDecompiler, we design decompilation-specific rewards that capture code validity, recompilability, execution consistency, and semantic fidelity. We further construct stage-aware diagnostic feedback from compiler errors, execution failures, and failed test cases, and introduce progress-aware trajectory rewarding to encourage beneficial revisions while suppressing regressions across refinement turns.

We train the AutoDecompiler family and conduct comprehensive evaluations across input settings, model scales, and benchmarks. Experimental results show that AutoDecompiler consistently outperforms its single-turn counterparts under the same model size and input setting, achieving clear improvements in behavioral re-executability (average {15.14\%} for pseudo-code input, and {14.65\%} for assembly input over the best 6.7B specialized LLMs). These results demonstrate that learning to exploit program feedback with reinforcement learning is an effective direction for improving the behavioral correctness of LLM-based binary decompilation.

\end{abstract}
\section{Introduction}
\label{intros}
Binary decompilation aims to recover high-level, C-like source-code representations from low-level executable binaries, enabling analysts to reason about program logic beyond machine instructions\cite{Jakub2017retdec}. 
It plays a critical role in software security analysis, vulnerability discovery, malware detection, and software maintenance, since most real-world software systems are released only in executable form and their original source code is often unavailable\cite{Wong2025DecLLM,Liu2025TheCS}. 
Therefore, practical decompilation requires not only syntactically readable outputs, but also semantic and behavioral consistency with the original binary.

Recent advances in large language models (LLMs) have enabled new possibilities for binary decompilation by leveraging their strong capabilities in code generation and semantic reasoning.
In general, existing LLM-based approaches can be categorized into two paradigms: end-to-end decompilation \cite{Nan2024Nova,feng2024sc2dec,tan2024LLM4Decompile,Liu2025TheCS} and pseudo-code based decompilation \cite{Wong2025DecLLM,hu2024degpt,she2024WaDec,Tan2025SK2DecompileLT,zhou2025fidelitygpt}.
End-to-end methods frame decompilation as a translation task, converting a sequence of assembly instructions into high-level source code in a single step by leveraging either commercial or fine-tuned LLMs.
Pseudo-code based methods take the pseudo-code representation produced by traditional decompilers as input, leveraging LLMs to analyze and optimize the pseudo-code. 

\begin{figure}
  \centering
  \includegraphics[height=1.0in,width=2.9in]{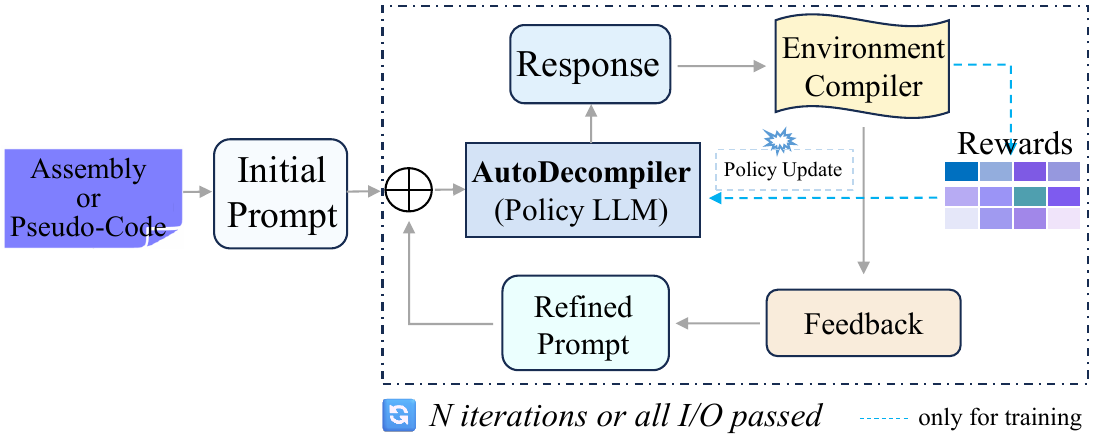}
  \caption{Overview of feedback-driven multi-turn binary decompilation.}
  \label{introoverview}
  \vspace{-3ex}
\end{figure}

Despite impressive progress, \textbf{existing LLM-based decompilation methods are still largely dominated by a single-turn generation paradigm, where LLMs reconstruct complete source code from fixed inputs in one pass}.
In practice, generated code can be validated through compilation, execution, and input/output (I/O) testing, producing diagnostic signals such as compiler errors, runtime exceptions, and failed I/O cases. These signals provide actionable guidance for locating, characterizing, and correcting errors in the generated code, enabling more reliable program recovery. However, current single-turn methods terminate after generation and therefore fail to leverage such feedback for iterative correction and progressive refinement.

These observations highlight the necessity of building decompilation LLMs that can dynamically diagnose, revise, and continuously improve their outputs through validation feedback.
\textbf{This raises a key research question: how can we transform LLM-based decompilation from a single-turn generation task into a feedback-driven multi-turn refinement process?}


A straightforward approach to enabling such iterative refinement is supervised fine-tuning (SFT). However, constructing high-quality supervision data that captures multiple rounds of execution and feedback is costly and difficult, while SFT primarily learns to imitate predefined refinement trajectories, limiting its ability to adaptively explore and improve through environment interactions. In contrast, reinforcement learning (RL) \cite{Wei2025ReinforcingMR,kumar2025iclr-training,jain2025iclrw-multiturn,wang2025deeptrans} provides a natural framework for this problem, as it directly optimizes policy models through feedback obtained from sequential interactions with the environment. This makes RL particularly suitable for learning multi-turn refinement behaviors, where each revision depends on prior outputs and diagnostic signals.

Motivated by this insight, we propose \textbf{AutoDecompiler}, \textit{a reinforcement-learning-trained LLM for feedback-driven multi-turn binary decompilation} (as shown in Figure~\ref{introoverview}). AutoDecompiler iteratively refines an initial decompiled-code candidate by leveraging compilation, execution, and testing feedback to progressively improve semantic and functional correctness.

However, building AutoDecompiler is non-trivial and faces several important challenges:

\noindent\textbf{Challenge 1: Defining reliable rewards for decompilation quality.}
Different from text generation tasks, the quality of decompiled code cannot be measured by a single objective.
A desirable decompilation output should be syntactically valid, semantically faithful to the original binary, structurally plausible as C code, and compilable/executable under I/O test cases.
Designing rewards that jointly capture these dimensions without encouraging superficial fixes is therefore non-trivial.

\noindent\textbf{Challenge 2: Converting heterogeneous feedback into effective refinement guidance.}
Validation feedback, such as compilation and execution results, provides useful diagnostic signals, but such feedback is often heterogeneous and weakly localized. Compiler errors, runtime failures, and failed I/O tests reveal different types of abnormal behaviors without directly indicating how to revise the code. Thus, a key challenge is how to transform such feedback into actionable guidance for subsequent refinement turns.

\noindent\textbf{Challenge 3: Ensuring progressive refinement across multiple turns.}
Multi-turn refinement does not naturally guarantee monotonic improvement. Although additional refinement turns provide opportunities to repair previous errors, they may also degrade previously correct behaviors or drift from the original program semantics when responding to local diagnostic feedback. In RL training, this issue is amplified because optimization may favor short-term reward gains while degrading overall decompilation quality.


To address the aforementioned challenges, we introduce several novel techniques into AutoDecompiler, each designed to tackle the corresponding challenge: 
\textbf{1) Weighted Multi-dimensional Decompilation Reward}.
We design a weighted reward that jointly evaluates C-like validity, compilation and execution correctness, syntactic consistency, and semantic fidelity. By integrating feedback from code validity checks, recompilation, I/O-based testing, and reference-guided structural-semantic comparison, this reward provides a more comprehensive and reliable optimization objective for learning decompilation refinement.
\textbf{2) Stage-aware Diagnostic Feedback Construction.}
We construct stage-aware diagnostic feedback to transform heterogeneous signals into actionable refinement guidance. After validating each generated code, the environment organizes failures by their stages, such as compilation, runtime, and I/O-output errors, and reformulates them into natural-language prompts for the next refinement turn. This enables the model to revise code based on concrete failure evidence rather than scalar rewards alone.
\textbf{3) Progress-aware Trajectory Rewarding.}
To prevent multi-turn refinement from drifting into arbitrary revisions, we incorporate a progress-aware trajectory reward. It jointly considers final and intermediate decompilation quality while explicitly penalizing regressions relative to the best previously achieved reward. This encourages the policy to make beneficial progress across refinement turns while avoiding revisions that degrade previously achieved results.

We train AutoDecompiler on the CID dataset~\cite{Liu2025TheCS}, a large-scale function-level decompilation dataset.
To support diverse decompilation scenarios, we train AutoDecompiler under two input settings: end-to-end decompilation (\textbf{AutoDecompiler-E2E}) and pseudo-code-based decompilation (\textbf{AutoDecompiler-Pscode}). For each setting, we further instantiate models at three parameter scales, including \textbf{1.3B, 6.7B, and 30B}, allowing users to trade off decompilation performance and computational cost. 
We evaluate the AutoDecompiler family on two widely used benchmarks, HumanEval~\cite{tan2024LLM4Decompile} and MBPP~\cite{austin2021mbpp},
using three complementary metrics: re-compilability, re-executability~\cite{Liu2025TheCS}, and code readability~\cite{Eom2024R2I}. Across matched settings of dataset, input representation, and model scale, AutoDecompiler consistently outperforms its single-turn counterparts, with particularly substantial gains in re-compilability and re-executability. 
It also achieves competitive performance against advanced coding agents and general-purpose LLMs, while maintaining favorable generalizability under cross-compiler and cross-decompiler settings.

\begin{figure*}
  \centering
  \includegraphics[height=1.2in,width=5.5in]{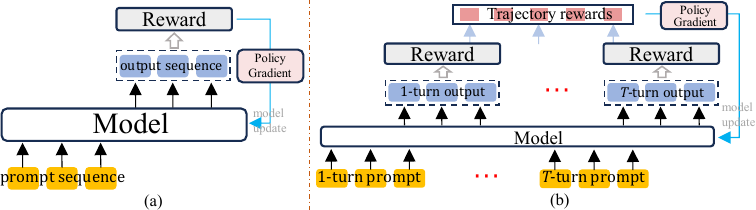}
  \caption{The core differentce between (a) single-turn RL and (b) our multi-turn RL.}
  \label{relatedwork}
  \vspace{-3ex}
\end{figure*}

In summary, this paper makes the following contributions:


\noindent\textbullet~
We formulate binary decompilation as a feedback-driven multi-turn refinement problem, and propose \textbf{AutoDecompiler}, a specialized LLM for function-level decompilation that is trained with reinforcement learning to iteratively refine generated decompiled code using validation feedback.

\noindent\textbullet~We design a decompilation-specific multi-dimensional reward for RL training. The reward jointly captures C-like validity, recompilability, re-executability, syntactic consistency, and semantic fidelity, providing a comprehensive optimization objective beyond rewards based solely on superficial code plausibility or compilation success.

\noindent\textbullet~We develop feedback- and turn-aware mechanisms for progressive multi-turn refinement. AutoDecompiler transforms compilation, execution, and I/O testing signals into stage-aware diagnostic feedback, while a progress-aware trajectory reward encourages beneficial revisions and penalizes regressions across refinement turns.

\noindent\textbullet~
We train the AutoDecompiler family and evaluate it across input settings, model scales, and benchmarks. Results demonstrate consistent improvements over single-turn counterparts, competitive performance against advanced coding agents and general-purpose LLMs, and favorable generalization across compilation and decompilation settings.

\noindent\textbullet~We open-source the AutoDecompiler family and associated resources to facilitate future research and practical applications at HuggingFace and Zenodo.


\section{Background \& Motivation}
\subsection{Reinforcement Learning}

Reinforcement learning is a learning paradigm that optimizes a policy through reward signals obtained from interactions with an environment. Representative RL methods include RLHF~\cite{Vemprala2023RLHF}, Proximal Policy Optimization (PPO)~\cite{Schulman2017PPO}, and recent critic-free variants such as GRPO~\cite{Shao2024GRPO} and REINFORCE++~\cite{jian2024REINFORCEpp}, which mainly differ in advantage estimation and token-level policy updates.

Due to its ability to model sequential decision-making and optimize non-differentiable objectives, RL has been widely explored for post-training large language models with human, verifiable, execution-based, or environmental feedback. CodeRL~\cite{le2022coderl}, O1-CODER~\cite{zhang2024o1coder}, and ACECODER~\cite{zeng2025acecoder} use actor-critic or Monte Carlo Tree Search\cite{wiechowski2021MonteCT}-enhanced RL to improve program synthesis. Reflect, Retry, Reward~\cite{Bensal2025ReflectRR} leverages policy-gradient RL with self-reflection and retry mechanisms to enhance multi-step reasoning. DeepTrans~\cite{wang2025deeptrans} applies GRPO to deep-reasoning translation, while Pentest-R1~\cite{kong2025pentestr1} combines offline demonstrations and online RL for autonomous penetration testing. 
WizardMath~\cite{ICLR2025WizardMath} introduces RLEIF to improve mathematical chain-of-thought reasoning through evolved instructions and process supervision.

Despite these advances, they optimize the generation of a single continuous token sequence, rewarding the model based on the quality of the sequence. In contrast, our work requires the model to perform iterative decompilation refinement through multiple rounds of sequence (i.e., decompiled code) generation, validation, feedback interpretation, and revision. Therefore, the optimization objective shifts from single-turn sequence generation to multi-turn feedback-driven sequence refinement. Figure~\ref{relatedwork} shows the difference between conventional single-turn RL and our multi-turn RL approach.

\begin{figure*}
  \centering
  \includegraphics[height=1.90in,width=5.7in]{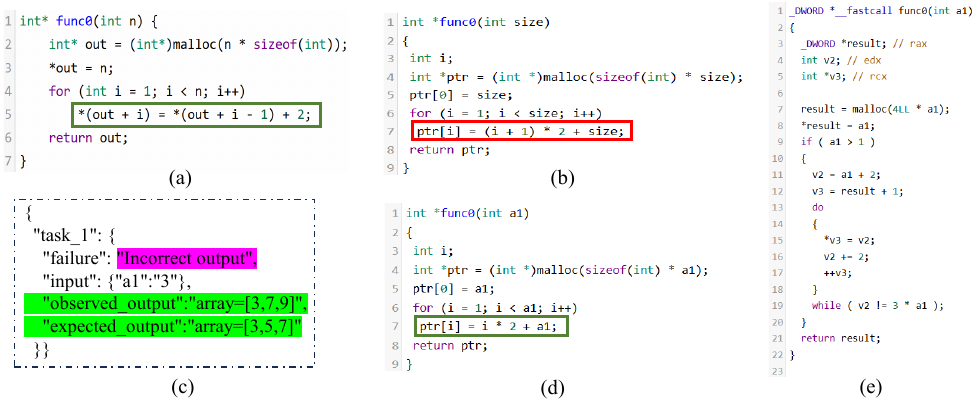}
\caption{Motivating example. Presented are (a) the ground-truth source code, (e) the IDA Pro pseudo-code, (b) the first-turn decompilation by LLM based the pseudo-code, (c) the validation feedback collected from testing (b), and (d) the second-turn decompilation by LLM after incorporating the feedback.}
\label{fig:motivation}
\vspace{-3ex}
\end{figure*}

\subsection{Decompilation with LLM}

In recent years, the success of LLMs has rapidly advanced decompilation technology. LLM-based decompilation
mainly follows two paradigms: end-to-end decompilation and pseudo-code-based decompilation. Representative end-to-end methods include LLM4Decompile-End~\cite{tan2024LLM4Decompile}, Nova~\cite{Nan2024Nova}, WaDec~\cite{she2024WaDec}, SALT4Decompile~\cite{Wang2025SALT4DecompileIS}, FAE~\cite{feng2024sc2dec}, and CIM~\cite{Liu2025TheCS}, which treat decompilation as a code translation task, training specialized LLMs to directly convert assembly code into C-like code. 

Pseudo-code-based methods include LLM4Decompile-Ref~\cite{tan2024LLM4Decompile}, SK2Decompile~\cite{Tan2025SK2DecompileLT}, D-LIFT~\cite{Zou2025DLiFTIL}, DecGPT~\cite{hu2024degpt}, FidelityGPT~\cite{zhou2025fidelitygpt}, and DecLLM~\cite{Wong2025DecLLM}. LLM4Decompile-Ref and SK2Decompile optimize pseudo-code from traditional decompiler via supervised training to produce more executable C code. D-LIFT designs the D-SCORE reward to drive GRPO for optimizing pseudo-code quality. FidelityGPT addresses fidelity issues in pseudo-code using retrieval-augmented generation, while DecGPT uses commercial LLMs to improve pseudo-code readability. 
DecLLM refines the pseudo-code through an iterative LLM repair loop with static compiler diagnostics and dynamic runtime feedback to produce recompilable and executable C code.

DecLLM~\cite{Wong2025DecLLM} is the closest work to ours, as it uses LLM to analyze execution feedback to optimize pseudo-code. However, DecLLM relies on a prompt-based repair loop with closed commercial general LLMs, which can raise confidentiality concerns for proprietary binaries and incur substantial API cost and latency. Moreover, its repair process is performed only at inference time and may suffer from unstable edits, hallucinations, or shortcut fixes. In contrast, our work trains a locally deployable, decompilation-specialized LLM that learns a feedback-driven multi-turn refinement policy, enabling more controllable and scalable decompilation refinement.

\subsection{Motivation}

Figure~\ref{fig:motivation} presents a motivating example for feedback-driven multi-turn decompilation. The ground-truth function in Figure~\ref{fig:motivation}(a) allocates an integer array, initializes the first element with the input size, and fills the remaining elements by increasing the value by two. Although the IDA Pro~\cite{hex-rays2025} pseudo-code in Figure~\ref{fig:motivation}(e) does not preserve the original source-level structure, it still exposes the key computation pattern.

However, the first-turn LLM decompilation in Figure~\ref{fig:motivation}(b) introduces a subtle semantic error. The generated code is readable and compilable, but the expression \texttt{(i + 1) * 2 + size} produces incorrect values. For example, given input \texttt{3}, it outputs \texttt{[3,7,9]} instead of the expected \texttt{[3,5,7]}, as shown in Figure~\ref{fig:motivation}(c). This case shows that syntactic correctness and readability alone are insufficient for reliable decompilation: a plausible-looking program may still deviate from the original binary behavior.

Execution feedback provides a concrete signal for diagnosing such hidden semantic errors. Guided by the failing test outputs, the model revises the loop assignment in the second turn, replacing \texttt{(i + 1) * 2 + size} with \texttt{i * 2 + a1}, as shown in Figure~\ref{fig:motivation}(d). This correction recovers the intended arithmetic sequence and aligns the generated code with the source-code behavior.

This example motivates our work. Instead of treating decompilation as a one-shot generation task, LLM-based decompilation can leverage validation feedback to iteratively diagnose and refine generated code. We therefore formulate decompilation as a feedback-driven multi-turn refinement problem and train the model with RL, enabling it to learn effective revision behaviors from feedback.

\section{Method}
\label{sec:framework}

\begin{figure*}
  \centering
  \includegraphics[height=1.95in,width=6.2in]{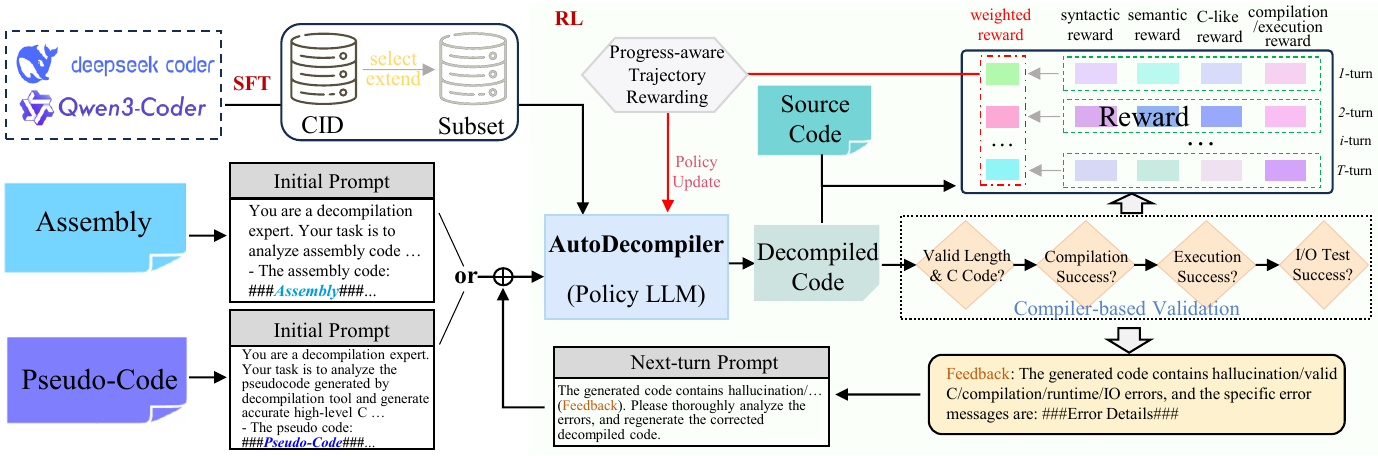}
  \caption{The details of building AutoDecompiler with SFT and RL.}
  \label{methodoverview}
  \vspace{-3ex}
\end{figure*}

This section presents the procedure of building AutoDecompiler (as shown at Figure~\ref{methodoverview}). Following the common post-training pipeline of domain-specialized LLMs~\cite{Hui2024Qwen25CoderTR,Guo2024DeepSeekCoderWT}, building AutoDecompiler consists of two training stages. The first stage performs domain-specialized SFT on the decompilation-specific CID~\cite{Liu2025TheCS} to adapt the base LLM to binary decompilation. This stage teaches the model to map low-level program representations, including assembly instructions or pseudo-code, into C-like source code. However, SFT only learns a static input-output mapping and does not explicitly teach the model how to diagnose and repair its own decompilation errors. Therefore, in the second stage, the SFT-initialized model is further optimized with RL to acquire feedback-driven multi-turn refinement capabilities. 
Specifically, at each refinement turn, the code candidate generated by the model is validated through C-code validity analysis, re-compilation, re-execution, and input/output testing. Across a refinement trajectory, the resulting validation signals serve two roles: scalar rewards are computed for multi-turn RL optimization, while stage-aware diagnostic feedback is concatenated to the next-turn prompt to guide subsequent refinement.

The rest of this section is organized as follows. Section~\ref{sec:formulation} defines the state, action, feedback, reward, and trajectory used in RL training. Section~\ref{sec:sft} introduces domain-specialized SFT for decompilation LLM initialization. Section~\ref{sec:rltraining} presents the key designs of the RL training stage that enable AutoDecompiler to achieve feedback-driven multi-turn refinement.


\subsection{RL Problem Formulation}
\label{sec:formulation}
We formulate feedback-driven multi-turn binary decompilation as an RL task. Let $b$ denote a binary function, $x$ denote the model input, and $\pi_\theta$ denote the policy model parameterized by $\theta$. Depending on the input setting, $x$ can be either the assembly representation of $b$ or the pseudo-code produced by a traditional decompiler. At refinement turn $t$, the model observes a state $s_t = (x, y_{<t}, f_{<t})$, where $y_{<t}$ denotes previous decompilation outputs and $f_{<t}$ denotes previous diagnostic feedback. The model then generates a complete C-like decompiled function $y_t \sim \pi_\theta(\cdot \mid s_t)$, which is treated as the action at turn $t$. We then validate $y_t$ through C-code checking, re-compilation, re-execution, and input/output testing, and obtain diagnostic feedback $f_t$ together with a reward score ${r}_t$.

A complete refinement trajectory during training is denoted as
$\tau = (s_1, y_1, f_1, {r}_1, \ldots, s_T, y_T, f_T, {r}_T)$, where $T<T_{max}$ is the number of refinement turns and $T_{max}$ denotes the maximum refinement budget. Our goal is to optimize the policy model $\pi_\theta$ to maximize the final decompilation quality while encouraging effective improvements and discouraging regressions across turns. After reinforcement learning, the optimized policy model $\pi_\theta$ serves as the final AutoDecompiler.

\subsection{Domain-Specialized SFT}
\label{sec:sft}
Before reinforcement learning, we first perform SFT to adapt the base LLM to the syntax, structure, and output style of binary decompilation. This initialization allows the subsequent RL stage to focus on learning feedback-driven refinement rather than learning decompilation from scratch.

\subsubsection{SFT Dataset}
\label{datasetbuilding}

We construct both the SFT and RL datasets from CID~\cite{Liu2025TheCS}, a large-scale decompilation corpus.
Since samples with no more than 10 CFG nodes account for about 95\% of CID, random sampling would overrepresent simple  samples. To obtain an RL dataset with broader complexity coverage, we use CFG-node count and instruction-token count as complementary lightweight complexity indicators. Specifically, we partition all samples into ten CFG-node groups (1--9 and $\geq$10 nodes) and select the 1,000 samples with the largest instruction-token counts from each group, yielding a 10K RL dataset. All remaining samples are used for SFT. For each sample, we retain its assembly, corresponding IDA Pro~\cite{hex-rays2025} pseudo-code, and reference source code.


We construct two types of SFT samples. For end-to-end decompilation, we follow the CIW~\cite{Liu2025TheCS} prompt format and use assembly as input and reference source code as the target. For pseudo-code-based decompilation, we learn the CIW-style instruction format, using decompiler-generated pseudo-code as input and the same reference source code as the target. The detailed pseudo-code-based prompt templates are provided in Appendix~\ref{appendsft}.

\subsubsection{SFT Training}
We train the SFT model using an instruction-conditioned autoregressive objective. Each training sample is formatted as a decompilation instruction, where the input part contains either assembly code or decompiler-generated pseudo-code, and the output part contains the reference  source code. The model is trained to generate the target  code token by token.

Formally, let $\mathbf{x}_{\mathrm{in}}=[x_1,\ldots,x_m]$ denote the input prompt and let $\mathbf{x}_{\mathrm{out}}=[x_{m+1},\ldots,x_n]$ denote the target source-code sequence. Given the concatenated sequence $\mathbf{x}=[\mathbf{x}_{\mathrm{in}},\mathbf{x}_{\mathrm{out}}]$, the model predicts each target token autoregressively as
\begin{equation}
P(x_i \mid x_{<i};\phi)=\mathrm{softmax}(\mathbf{W}_o \mathbf{h}_i)_{x_i}
\end{equation}
where $i\in {(m+1,...,n)}$, $\mathbf{h}_i$ is the hidden state at position $i$, $\mathbf{W}_o$ is the output projection matrix, and $\phi$ denotes the trainable parameters of SFT.

The SFT objective minimizes the negative log-likelihood of the target source-code tokens:
\begin{equation}
\mathcal{L}_{\mathrm{SFT}}(\phi)=
-\sum_{i=m+1}^{n}
\log P(x_i \mid x_{<i};\phi)
\end{equation}


\subsection{RL for AutoDecompiler}
\label{sec:rltraining}
Unlike SFT, which mainly teaches the model to imitate static input-output pairs, RL enables the model to optimize non-differentiable validation objectives and learn refinement behaviors from validation feedback. This capability is essential for feedback-driven decompilation, where the model is expected to revise previous generated code according to such feedback.

This subsection first introduces the RL dataset and then presents the key training designs, including multi-dimensional reward design, stage-aware diagnostic feedback, progress-aware trajectory rewarding, and the final policy optimization objective. 
Considering GRPO’s~\cite{Shao2024GRPO} efficiency and critic-free optimization, we adopt it as the base RL algorithm to optimize feedback-driven multi-turn refinement trajectories.

\subsubsection{RL Dataset}
As described in Section~\ref{datasetbuilding}, we use only 10K samples for RL training due to the high computational cost of multi-turn rollout and program validation. The RL dataset is strictly disjoint from the SFT dataset to avoid data leakage between supervised initialization and reinforcement learning.

Each RL sample keeps the same basic content as the SFT samples. In addition, to enable validation of generated decompiled code during RL training, we augment each sample with dependency information and input/output test cases. The dependency information is used to construct the re-compilation context, while the input/output cases are used to assess the execution behavior of generated code. Thus, unlike SFT samples that provide static input-output supervision, RL samples provide the validation resources needed to produce diagnostic feedback and reward signals during multi-turn refinement. The core elements of an RL training example are presented in Appendix~\ref{appendrl}.

\subsubsection{Weighted Multi-dimensional Decompilation Reward}
\label{sec:reward}
We design a multi-dimensional reward to evaluate each generated decompilation, consisting of the following components:

\textbf{C-like validity.}
To prevent hallucinated or malformed outputs, we introduce a C-like validity reward that evaluates whether the model output is a valid C-like decompiled function. 

Specifically, we examine whether the output contains a complete function body, follows C-like syntax, and avoids non-code content such as explanations, Markdown formatting, incomplete fragments, or repetitive hallucinated text. This reward also serves as the first-stage gate for subsequent validation: if the output cannot be recognized as valid C-like code, it receives a negative validity reward, 
while the remaining validation stages and reward computations are skipped to avoid unnecessary validation and computation overhead.
Formally, the reward is defined as
\begin{equation}
R_{\mathrm{C}}(y_t)=
\begin{cases}
1, & \text{if } y_t \text{ is a valid C-like function}, \\
-1, & \text{otherwise}
\end{cases}
\label{eq:c_reward}
\end{equation}





\textbf{Compilation and execution correctness.} This reward measures whether the decompiled code can be re-compiled and re-executed, and whether its execution behavior is consistent with the expected behavior. It provides the main verifiable signal.

After a decompiled code candidate passes the C-like validity check, we first place it into the corresponding re-compilation context and invoke the compiler. If re-compilation fails, the candidate receives the lowest reward because execution-based validation cannot proceed. If re-compilation succeeds, we execute the re-compiled program with the provided test inputs. 
A runtime error, such as abnormal termination or timeout, receives a negative reward that is less severe than the penalty for re-compilation failure, but still indicates invalid execution behavior.
If the program executes successfully but produces outputs inconsistent with the expected outputs, it receives a partial penalty for being behaviorally incorrect. Only candidates that successfully re-compile and pass all input/output tests receive the highest reward.

Formally, the compilation and execution correctness reward is defined as:
\begin{equation}
R_{\mathrm{exe}}(y_t)=
\begin{cases}
-1, & \text{re-compilation fails},\\
-0.6, & \text{runtime error},\\
-0.3, & \text{I/O mismatch},\\
1, & \text{all I/O tests pass}
\end{cases}
\end{equation}

This staged reward provides a graded optimization signal: it encourages the model to first generate compilable code, then executable code, and finally behaviorally consistent code. Compared with a binary pass/fail reward, this design provides denser feedback for RL training and better supports progressive refinement across turns. In our setting, this reward is particularly important because a decompiled function may be readable and compilable while still deviating from the behavior of the original binary.

\textbf{Syntactic reward.} 
The compilation and execution reward provides direct feedback on re-compilability and runtime behavior, but it is still relatively coarse and may not sufficiently guide the structural form of decompiled code. To provide a denser structural signal, we introduce a syntactic reward that measures AST-level similarity between the generated decompiled code and the reference source code. This reward encourages the model to preserve source-like syntactic structures, such as function bodies, branches, loops, assignments, and expressions, rather than only producing code that passes compilation or tests.

Specifically, we parse both the decompiled code $y_t$ and the reference code $y^\star$ using Tree-sitter, remove comments, and extract non-leaf AST subtrees from each parsed tree. The syntactic reward is computed as the fraction of reference subtrees that can be matched in the decompiled code:
\begin{equation}
R_{\mathrm{syn}}(y_t,y^\star)=
\frac{
\sum_{s\in \mathrm{ASTSub}(y^\star)}
\mathbb{I}[s\in \mathrm{ASTSub}(y_t)]
}{
|\mathrm{ASTSub}(y^\star)|
}
\end{equation}
Here, $\mathrm{ASTSub}(\cdot)$ denotes the collection of extracted non-leaf AST subtrees, including repeated subtree instances, and $\mathbb{I}[\cdot]$ is the indicator function. If parsing fails or no valid reference subtree can be extracted, the syntactic reward is set to $0$. Compared with token-level matching, 
this AST-based reward captures richer syntactic structure and can better reflect errors such as incorrect control-flow nesting, missing operators, or malformed expressions.

\textbf{Semantic reward.}
Although syntactic reward captures structural similarity, two programs with similar AST structures may still differ in how values are propagated among variables. To provide a more semantics-aware training signal, we introduce a semantic reward based on data-flow matching. The intuition is that decompiled code should not only resemble the reference code syntactically, but should also preserve key variable dependencies and value-propagation relations. 

Specifically, we parse both the decompiled code $y_t$ and the reference code $y^\star$ using Tree-sitter and extract data-flow relations while handling declarations, assignments, increments, and control-flow statements. Each data-flow item records a normalized target variable, its dependency relation, and the normalized source variables from which its value is derived. To reduce the influence of superficial naming differences, 
variable names are normalized into abstract identifiers during data-flow normalization.
The semantic reward is then computed as the fraction of reference data-flow relations that can be matched in the decompiled code:
\begin{equation}
R_{\mathrm{sem}}(y_t,y^\star)=
\frac{
M(\mathrm{DFG}(y^\star),\mathrm{DFG}(y_t))
}{
|\mathrm{DFG}(y^\star)|
}
\end{equation}
Here, $\mathrm{DFG}(\cdot)$ denotes the normalized data-flow relations extracted from code, and ${M}(\cdot,\cdot)$ is the number of one-to-one matched relations. If no valid data-flow relation can be extracted from the reference code or parsing fails, the semantic reward is set to $0$. 

This reward encourages the model to preserve variable-use relations and value dependencies, such as whether an assigned variable is computed from the correct operands or whether loop-updated variables preserve the expected dependency relations.
Therefore, it provides complementary guidance to the compilation, execution, and syntactic rewards.

\begin{table*}[h!]
\caption{Details of constructing the stage-aware
diagnostic feedback and next-turn prompt. \texttt{\#<Error Details>\#} is the failure information captured from environtment compiler.}
\centering
    \resizebox{\textwidth}{!}{
{
   \fontsize{8pt}{7pt}\selectfont
\begin{tabular}{|c|l|l|l|}
\hline
\multicolumn{1}{|l|}{Priority} & Failure Category & Stage-aware  Diagnostic  Feedback ($f_t$) & Next-turn Prompt \\ \hline
1   & Excessive Length               & “Code is excessively long, indicating repetitive hallucination.”                                &$f_t$+“Please regenerate the code.”               \\ \hline
2   & Invalid Code               &   “The generated code is not a C language program.                             ”&$f_t$+“Please regenerate the code.”\\ \hline
3   & Compilation Error              & \begin{tabular}[c]{@{}l@{}}“The generated code contains compilation errors, and the \\ specific error messages are: \texttt{\#<Error Details>\#}.”\end{tabular}                     &$f_t$+\begin{tabular}[c]{@{}l@{}}“Please analyze the errors \\and regenerate the code.”\end{tabular}    \\ \hline
4   & Runtime Error             & \begin{tabular}[c]{@{}l@{}}“The generated code contains runtime errors, and the specific \\ error messages are: \texttt{\#<Error Details>\#}.”\end{tabular}  &$f_t$+\begin{tabular}[c]{@{}l@{}}“Please analyze the errors \\and regenerate the code.”\end{tabular} \\ \hline
5   & I/O Mismatch               &\begin{tabular}[c]{@{}l@{}} “There are incorrect outputs when performing unit testing on \\ the generated code. Specifically, it is: \texttt{\#<Error Details>\#}.”\end{tabular}  &$f_t$+\begin{tabular}[c]{@{}l@{}}“Please analyze the errors \\and regenerate the code.”\end{tabular} \\ \hline
6   & All right               & “The generated decompiled is OK.”&$f_t$+ “Stop generating.”\\ \hline
\end{tabular}
}}
\label{feedbackbuidling}
\vspace{-3ex}
\end{table*}

The reward score at turn $t$ is computed as a weighted sum:
\begin{equation}
{r}_t
=
w_C R_{\mathrm{C}}
+
w_{\mathrm{exe}} R_{\mathrm{exe}}
+
w_{\mathrm{syn}} R_{\mathrm{syn}}
+
w_{\mathrm{sem}} R_{\mathrm{sem}}
\end{equation}Here, $w_C$, $w_{\mathrm{exe}}$, $w_{\mathrm{syn}}$, and $w_{\mathrm{sem}}$ are reward weights. For outputs that fail the C-like validity check, we set $R_{\mathrm{exe}}=-1$, $R_{\mathrm{syn}}=R_{\mathrm{sem}}=0$. This scalar reward is subsequently used as the turn-level quality score for progress-aware trajectory rewarding and GRPO optimization. In our implementation, the weights are chosen to keep execution-oriented correctness as the primary optimization signal, while syntactic and data-flow rewards
provide auxiliary source-oriented shaping signals.

\subsubsection{Stage-aware Diagnostic Feedback Construction}
\label{sec:environment}

During reward computation, we record the validation stage and validation-failure information for each generated code candidate, together with the resulting reward outcome. AutoDecompiler leverages these results to construct stage-aware diagnostic feedback. Specifically, AutoDecompiler first identifies the failure category according to the validation stage and the corresponding reward outcome. It then converts the failure category into a natural-language diagnostic message and embeds the raw validation-failure information as concrete details. The resulting diagnostic feedback is concatenated with the next-turn prompt, so that the model can refine its previous output based on explicit validation evidence.

We instantiate this construction for different validation outcomes as follows: (1) If the generated code is excessively longer than the reference code, the failure category is defined as repetitive hallucination, and the diagnostic message explains that the code should be regenerated in a concise form. (2) If the output cannot be recognized as a C-like program, the failure category is invalid code, and the message states that the generated result is not valid C-like code. (3) If re-compilation fails, the failure category is compilation error, and the compiler diagnostics are serialized and embedded as concrete error details, such as syntax errors, undeclared identifiers, type mismatches, or missing declarations. (4) If re-compilation succeeds but execution fails, the failure category is runtime error, and the corresponding runtime information, such as abnormal termination or timeout, is included as validation-failure details. (5) If execution succeeds but the produced outputs are inconsistent with the expected outputs, the failure category is input/output mismatch, and the failed test cases are embedded into the diagnostic message. (6) When all tests pass, the validation returns a success signal and the refinement trajectory terminates successfully; otherwise, the constructed diagnostic feedback is concatenated to the next-turn prompt, and the refinement process continues until the maximum refinement budget is reached.

Table~\ref{feedbackbuidling} shows the detailed content of the stage-aware diagnostic feedback and next-turn prompt. The entire refinement process is performed automatically.

\subsubsection{Progress-aware Trajectory Rewarding}
\label{sec:mtgrpo}

In feedback-driven multi-turn decompilation, using only the final-turn reward cannot fully characterize whether the model has learned meaningful refinement behavior. A trajectory may obtain a high final reward even if some intermediate turns introduce regressions or produce low-quality outputs. Therefore, we introduce a progress-aware trajectory rewarding mechanism that jointly considers final and intermediate decompilation quality while explicitly penalizing regressions relative to the best previously achieved reward.

For a refinement trajectory $\tau$, let $r_t$ denote the weighted turn reward obtained after refinement turn $t$, and let $T$ denote the number of valid refinement turns. 
First, we use the reward of the final valid turn as the primary quality objective:
\begin{equation}
R^{\mathrm{final}} = r_T
\end{equation}

Since the initial generation provides the starting point for subsequent refinement, we further introduce a first-turn quality term:
\begin{equation}
R^{\mathrm{first}} = r_1
\end{equation}

To preserve the quality of intermediate refinement states, we additionally compute the historical refinement quality:
\begin{equation}
R^{\mathrm{hist}}= \begin{cases} 0, & T=1,\\ \frac{1}{T-1}\sum_{t=1}^{T-1}r_t, & T>1. \end{cases}
\end{equation}
The historical term encourages the model to maintain useful intermediate solutions instead of relying solely on late-stage corrections.

To explicitly penalize regressions during refinement, we track the best reward achieved before each turn and define the regression penalty:
\begin{equation}
    r^{\mathrm{best}}_{t-1}=\max(r_1, ..., r_{t-1})
\end{equation}
\begin{equation}
    R^{\mathrm{reg}}= \begin{cases} 0, & T=1,\\ \frac{1}{T-1}\sum_{t=2}^{T} [r^{\mathrm{best}}_{t-1}-r_t-\epsilon]_+, & T>1. \end{cases}
\end{equation}
where $[x]_+=\max(x,0)$ and $\epsilon$ is a tolerance margin. This term penalizes refinement steps that degrade previously achieved quality, while tolerating small reward fluctuations.

The final trajectory reward is defined as:
\begin{equation}
    R^{\mathrm{traj}}=R^{\mathrm{final}}+\lambda_f R^{\mathrm{first}}+\lambda_h R^{\mathrm{hist}}-\beta R^{\mathrm{reg}}
\end{equation}
where $\lambda_f$=0.05, $\lambda_h$=0.15, $\beta$=0.20 in our implementation. 

This formulation preserves final decompilation quality as the primary optimization objective while encouraging high-quality intermediate refinements and discouraging destructive revisions.

\subsubsection{Policy Optimization Objective}
\label{sec:policyopt}

After obtaining the trajectory reward, 
we follow GRPO~\cite{Shao2024GRPO} for policy optimization. 
For each input $x$, the old policy model samples a group of $G$ trajectories $\{\tau_i\}_{i=1}^{G}$, where $\tau_i \sim \pi_{\theta_{\mathrm{old}}}(\cdot\mid x)$. Given their trajectory rewards $\{R_i^{\mathrm{traj}}\}_{i=1}^{G}$, the group-relative advantage is computed as:
\begin{equation}
A_i=
\frac{
R_i^{\mathrm{traj}}-\mu_G
}{
\sigma_G+\epsilon
}
\end{equation}
where $\mu_G$ and $\sigma_G$ are the mean and standard deviation of the trajectory rewards within the group, and $\epsilon$ is a small constant for numerical stability. The same trajectory-level advantage is assigned to all model-generated tokens in $\tau_i$.

Let $a_{i,t,l}$ denote the $l$-th generated token at refinement turn $t$ of trajectory $\tau_i$. The probability ratio between the current and old policies is:
\begin{equation}
\omega_{i,t,l}(\theta)=
\frac{
\pi_{\theta}(a_{i,t,l}\mid c_{i,t,l})
}{
\pi_{\theta_{\mathrm{old}}}(a_{i,t,l}\mid c_{i,t,l})
}
\end{equation}
where $c_{i,t,l}$ denotes its autoregressive context. 

The policy is optimized using the clipped GRPO loss:
\begin{equation}
\scriptsize
\mathcal{L}_{\mathrm{GRPO}}(\theta)
=
-\mathbb{E}
\left[
\min
\left(
\omega_{i,t,l}(\theta)A_i,
\mathrm{clip}\!\left(
\omega_{i,t,l}(\theta),1-\epsilon_c,1+\epsilon_c
\right)A_i
\right)
\right]
\end{equation}
where $\epsilon_c$ is the clipping threshold. 

\section{Experiments}
\label{mainexpes}

In this section, we evaluate AutoDecompiler on two benchmark datasets.
The experiments are designed to answer the following research questions:
\begin{itemize}
    \item \textbf{RQ1:} How does AutoDecompiler compare with traditional decompilers, state-of-the-art (SOTA) decompilation LLMs, advanced general-purpose LLMs and coding agents?
    \item \textbf{RQ2:} How effectively does AutoDecompiler leverage multi-turn reinforcement learning to progressively improve decompilation performance?
    \item \textbf{RQ3:} How does each key component contribute to AutoDecompiler?
    \item \textbf{RQ4:} How well does AutoDecompiler generalize across different compilation and decompilation toolchains?
    \item \textbf{RQ5:} How does AutoDecompiler repair decompilation errors in practical cases?
\end{itemize}

\subsection{Experimental Setups}
\label{mainexesetups}

\subsubsection{Evaluation Benchmarks}
Following prior works~\cite{tan2024LLM4Decompile,feng2024sc2dec,Nan2024Nova,Liu2025TheCS,tan2025decombench,Tan2025SK2DecompileLT,Wang2025SALT4DecompileIS}, we use \textbf{\textit{HumanEval}}~\cite{chen2021codex} and \textbf{\textit{MBPP}}~\cite{austin2021mbpp} as evaluation benchmarks. HumanEval and MBPP are originally Python programming benchmarks containing 164 and 974 tasks, respectively. We use their C-language adaptations provided by~\cite{tan2025decombench}, in which the Python solutions and test cases are manually translated into C for decompilation evaluation.


\subsubsection{Data Leakage Control}
\label{dataleakacontrol}
To prevent training-test contamination, 
we perform source-level decontamination of CID against the HumanEval/MBPP evaluation sets
before constructing the SFT and RL datasets. Following the procedure of CID~\cite{Liu2025TheCS}, we remove exact duplicates using normalized source-code hashes and structural duplicates using normalized AST hashes. CID has already been decontaminated against HumanEval, while we apply the same procedure to the MBPP benchmark. If a source function overlaps with an evaluation function, all of its compiled variants are excluded from training. Since these benchmarks are publicly available, potential exposure from the base LLMs' pre-training corpus cannot be fully excluded; our decontamination therefore prevents additional benchmark leakage during SFT and RL.

\subsubsection{Evaluation Metrics}
Drawing on prior work~\cite{tan2024LLM4Decompile,Tan2025SK2DecompileLT,Liu2025TheCS}, we adopt the following three metrics:

\textbf{Re-compilation Rate (Re-com)} measures whether the generated decompiled code can be compiled into executable binaries without compilation errors. A higher re-compilation rate suggests that the generated code better satisfies the syntactic and type constraints of the target language.

\textbf{{Re-executability Rate} (Re-exe)} assesses whether the decompiled code can run correctly and produce outputs consistent with the expected results. A higher Re-exe indicates greater \textbf{behavioral consistency} with the original program under the evaluated test cases.

\textbf{{Relative Readability Index} (R2I)}~\cite{Eom2024R2I} evaluates the readability of generated decompiled code in a relative manner. It extracts readability-related features from the abstract syntax tree, covering code quality, user preferences, conflicting features, syntactic errors, and general code characteristics, and aggregates them using predefined weights into a score between 0 and 1. A higher R2I score indicates better readability among the compared outputs.

\subsubsection{Baselines}

We compare AutoDecompiler with two traditional decompilers, Ghidra~\cite{Ghidra2025} and IDA Pro~\cite{hex-rays2025}, as well as representative LLM-based decompilation methods. We evaluate these LLM-based baselines under two input settings: end-to-end decompilation and pseudo-code-based decompilation.

For general-purpose LLMs, we consider two advanced models, GPT-5.6~\cite{openai2026gpt56} and DeepSeek-V4~\cite{deepseekai2026deepseekv4}, together with their corresponding coding agents, Codex (GPT-5.6)~\cite{chen2021codex} and DeepSeek Harness (DeepSeek-V4)~\cite{deepseekharness2026}. 
Due to the high computational cost and latency of evaluating general-purpose LLMs and agents, we randomly assign the two backbone LLMs to the two input settings (one per setting) and evaluate each LLM and its corresponding coding agent.
For pseudo-code-based decompilation, we additionally evaluate the DecLLM~\cite{Wong2025DecLLM} variant built on the same selected LLM.

For decompilation-specialized LLMs, we select several recent state-of-the-art works: FAE~\cite{feng2024sc2dec}, LLM4Decompile-DCBench~\cite{tan2025decombench} and CIM~\cite{Liu2025TheCS} for end-to-end decompilation, and Idioms~\cite{Dramko2025IdiomsND} and SK2Decompile~\cite{Tan2025SK2DecompileLT} for pseudo-code-based decompilation. 

More details of the baseline LLMs are provided in Appendix~\ref{appbasellms}.

\subsubsection{Implementation}


All functions in the training and test sets are compiled with GCC under four optimization levels, ranging from \texttt{-O0} to \texttt{-O3}. The resulting binaries are then processed by IDA Pro\footnote{In this paper, we use IDA Pro to ensure a fair comparison with the baselines. In practical applications, other disassemblers and decompilers can also be used.}~\cite{hex-rays2025} to obtain x86\_64 assembly and initial pseudo-code, which serve as the inputs for end-to-end decompilation and pseudo-code-based decompilation methods, respectively.

We use LLaMA-Factory~\cite{zheng2024llamafactory} for SFT and TRL~\cite{vonwerra2020trl} for RL training. 
Following LLM4Decompile~\cite{tan2024LLM4Decompile}, which serves as the base model for FAE, LLM4Decompile-DCBench, CIM, and SK2Decompile, we adopt its underlying DeepSeek-Coder~\cite{guo2024deepseekcoder} as the base model for training the 1.3B and 6.7B variants of AutoDecompiler.
For the 30B variant, we use Qwen3-Coder~\cite{Cao2026Qwen3CoderNextTR} as the base model.

During SFT, we set the learning rate to $5\times10^{-5}$ and train for 1 epoch. During RL training, we set the initial learning rate to $1\times10^{-6}$ and train for 2 epochs. The maximum input length is set to 4096 tokens. 
Since each refinement turn retains the generated code and diagnostic feedback in the trajectory history, the context grows substantially with the number of turns. Given the 16K-token context limit of DeepSeek-Coder, we set \(T_{\max}=3\) as the maximum refinement budget during training and inference. Trajectories can terminate earlier once validation succeeds.

To determine the reward weights and evaluate the contribution of key design components, we conduct weight selection experiments and ablation studies using the 1.3B end-to-end model on HumanEval. The smaller model size reduces the training cost of repeated controlled experiments, while the relatively small HumanEval test set enables efficient evaluation.
The weights of the four reward components
are initially set to 1 and then adjusted through controlled experiments. Finally, we set $w_C=0.5$, $w_{\mathrm{exe}}=1.0$, $w_{\mathrm{syn}}=0.3$, and $w_{\mathrm{sem}}=0.2$. 

During inference, we use greedy decoding and employ vLLM~\cite{kwon2023vllm} to accelerate inference. All experiments are conducted on two GPU clusters, each equipped with 8 NVIDIA H100-80GB GPUs.

\subsection{Main Results (RQ1)}

As introduced in Section~\ref{intros}, we refer to AutoDecompiler with assembly instructions as input for end-to-end decompilation as \textbf{AutoDecompiler-E2E}, and the one with pseudo-code as input as \textbf{AutoDecompiler-Pscode}. Each variant is compared with baseline models under the same input setting to ensure a fair comparison.

\begin{table*}[h!]

\caption{Comparison of traditional decompilers and end-to-end decompilation methods. *: Results are from the original paper, except that the MBPP result on CIM-6.7B is reproduced by us. (\%)}
\centering
    \resizebox{\textwidth}{!}{
   \scriptsize
\begin{tabular}{c|ccccccccccc}
\toprule
 \multirow{2}{*}{Metric}&  \multirow{2}{*}{Model}&\multicolumn{5}{c}{HumanEval}&\multicolumn{5}{c}{MBPP}\\
\cmidrule(lr){3-7} \cmidrule(lr){8-12}
&  & O0 & O1 & O2 & O3 & AVG & O0 & O1 & O2 & O3 & AVG \\
\hline
\multirow{9}{*}{Re-com}
& IDA Pro & 26.22 & 21.34 & 26.83 & 25.61 & 25.00 & 14.52 & 10.55 & 9.67 & 10.07 & 11.20 \\
 & Ghidra & 21.95 & 19.51 & 15.85 & 15.24 & 18.14 & 17.40 & 16.87 & 12.24 & 13.55 & 15.02 \\
 & GPT-5.6 & 89.02 & 81.71 & 84.15 & 81.71 & 84.15 & 88.88 & 86.55 & 85.65 & \textbf{85.90} &\textbf{ 86.75} \\
 & Codex (GPT-5.6) & 91.34 & 85.85 & 79.76 & 70.61 & 81.89 &\textbf{ 90.24} & 76.61 & 68.19 & 54.43 & 72.37 \\
 & FAE-6.7B* & 92.07 & 93.29 & 92.07 & 93.90 & 92.84 & - & - & - & - & - \\
 & CIM-6.7B* & 93.29 & 91.46 & 93.29 & 92.07 & 92.53 & 85.68 & 83.47 & 83.76 & 79.66 & 83.14 \\
 & AutoDecompiler-E2E-1.3B & 94.50 & 85.71 & 86.14 & 85.88 & 88.06 & 87.01 & 83.46 & 81.64 & 77.78 & 82.47 \\
 & AutoDecompiler-E2E-6.7B & 98.43 & \textbf{98.66} & 96.62 & 95.62 & 97.33 & 87.93 & 85.42 & 83.67 & 81.47 & 84.62 \\
 & AutoDecompiler-E2E-30B & \textbf{99.21} & 97.97 & \textbf{97.28} & \textbf{95.68} & \textbf{97.54} & 89.63 &\textbf{ 87.87} &\textbf{ 85.38} & 81.01 & 85.97 \\
\midrule
\multirow{10}{*}{Re-exe} & IDA Pro & 23.17 & 17.07 & 21.95 & 20.73 & 20.73 & 11.23 & 8.48 & 8.16 & 8.24 & 9.03 \\
 & Ghidra & 19.51 & 14.63 & 12.20 & 11.59 & 14.48 & 13.59 & 10.80 & 6.19 & 6.59 & 9.29 \\
 & GPT-5.6 & 78.05 & 52.44 & 54.88 & 54.88 & 60.06 & 65.91 & 53.33 & 51.66 & 52.20 & 55.78 \\
 & Codex (GPT-5.6) & 64.51 & 72.44 & 66.95 & 61.46 & 66.34 & 60.58 & 55.76 & 49.31 & 41.79 & 51.86 \\
 & FAE-6.7B* & 71.95 & 53.66 & 48.78 & 45.73 & 55.03 & - & - & - & - & - \\
 &LLM4Decompile-DCBench-6.7B* &61.59& 30.18& 34.15& 32.01& 39.48 &58.32 &39.58& 39.73& 37.06& 43.67 \\
 & CIM-6.7B* & 80.49 & 57.93 & 56.71 & 53.05 & 62.05 & 67.03 & 53.44 & 50.43 & 46.57 & 54.37 \\
 & AutoDecompiler-E2E-1.3B & 86.24 & 69.39 & 65.35 & 71.76 & 73.19 & 70.13 & 51.17 & 48.08 & 46.34 & 53.93 \\
 & AutoDecompiler-E2E-6.7B & 91.34 & 77.85 & 73.65 & 74.45 & 79.32 & 75.86 & 65.89 & 64.00 & 59.86 & 66.40 \\
 & AutoDecompiler-E2E-30B & \textbf{94.49} & \textbf{85.81} & \textbf{87.76} & \textbf{84.17} &\textbf{ 88.06} & \textbf{78.67} & \textbf{70.93} & \textbf{64.95} & \textbf{63.46} &\textbf{ 69.50} \\
\midrule
\multirow{8}{*}{R2I} & IDA Pro & 40.25 & 37.72 & 34.35 & 34.43 & 36.69 & 52.47 & 38.93 & 37.70 & 37.73 & 41.71 \\
 & Ghidra & 43.32 & 38.49 & 39.05 & 37.70 & 39.64 & 39.49 & 35.23 & 33.53 & 34.08 & 35.58 \\
 & GPT-5.6 & 63.60 & 64.64 & 63.16 & 66.70 & 64.53 & \textbf{71.06} & 66.83 & 64.73 & 65.64 & 67.07 \\
 & Codex (GPT-5.6) & 63.81 & 59.41 & 59.78 & 58.39 & 60.35 & 72.12 & 59.64 & 58.40 & 58.21 & 62.09 \\
 & CIM-6.7B* & \textbf{67.94} & 69.94 & \textbf{71.34} & \textbf{72.18} & \textbf{70.35} & 69.80 & \textbf{72.40} & \textbf{70.60} & \textbf{69.50} & \textbf{70.58} \\
 & AutoDecompiler-E2E-1.3B & 63.50 & \textbf{71.75} & 70.30 & 70.48 & 69.01 & 68.65 & 69.43 & 68.42 & 67.19 & 68.42 \\
 & AutoDecompiler-E2E-6.7B & 65.88 & 69.34 & 70.15 & 71.05 & 69.10 & 68.83 & 70.41 & 68.74 & 66.92 & 68.72 \\
 & AutoDecompiler-E2E-30B & 64.85 & 70.83 & 69.99 & 70.46 & 69.03 & 68.88 & 71.11 & 70.16 & 68.16 & 69.58 \\
\bottomrule
\end{tabular}%
}
\label{e2e67result}
\end{table*}

\begin{table*}[h!]
\caption{Comparison of pseudo-code-based decompilation methods. *: Results are from the SK2Decompile paper~\cite{Tan2025SK2DecompileLT}. $^{\dagger}$: Our reproduction follows the settings in the original paper. (\%)}
\centering
    \resizebox{\textwidth}{!}{
   \scriptsize
\begin{tabular}{c|ccccccccccc}
\toprule
 \multirow{2}{*}{Metric}&  \multirow{2}{*}{Model}&\multicolumn{5}{c}{HumanEval}&\multicolumn{5}{c}{MBPP}\\
\cmidrule(lr){3-7} \cmidrule(lr){8-12}
&  & O0 & O1 & O2 & O3 & AVG & O0 & O1 & O2 & O3 & AVG \\
\hline
\multirow{6}{*}{Re-com} & DeepSeek-V4 & 77.44 & 76.22 & 82.32 & 78.05 & 78.51 & 74.25 & 79.15 & 84.59 & 80.22 & 79.55 \\
 & DeepSeek Harness (DeepSeek-V4) & 95.73 & 96.34 & 95.12 & 95.12 & 95.58 & 82.39 & 73.45 & 67.07 & 63.92 & 71.71 \\
 & DecLLM (DeepSeek-V4)$^{\dagger}$ & 98.17 & 95.12 & 93.90 & 92.68 & 94.97 & \textbf{91.45} & 88.00 & 85.65 & 82.78 & 86.97 \\
 & AutoDecompiler-Pscode-1.3B & \textbf{100.00} & 97.78 & 97.50 & 93.59 & 97.22 & 87.32 & 86.85 & 84.70 & 82.93 & 85.45 \\
 & AutoDecompiler-Pscode-6.7B & 97.71 & 95.37 & 96.04 & 92.00 & 95.28 & 90.19 & \textbf{89.90} &\textbf{ 87.58} & \textbf{84.99} & \textbf{88.17} \\
 & AutoDecompiler-Pscode-30B & 98.47 & \textbf{98.11} & \textbf{100.00} & \textbf{98.85} & \textbf{98.86} & 87.17 & 86.00 & 84.07 & 80.49 & 84.43 \\
\midrule
\multirow{8}{*}{Re-exe} & DeepSeek-V4 & 73.78 & 67.07 & 76.22 & 70.73 & 71.95 & 59.22 & 61.45 & 63.44 & 58.79 & 60.73 \\
 & DeepSeek Harness (DeepSeek-V4) & 95.73 & 94.51 & 92.07 & 92.68 & 93.75 & 69.21 & 63.39 & 56.65 & 54.21 & 60.87 \\
 & DecLLM (DeepSeek-V4)$^{\dagger}$ & 90.85 & 84.76 & 80.49 & 78.05 & 83.54 & 77.34 & 70.06 & 64.50 & 57.69 & 67.40 \\
 & Idioms* & 70.73 & 25.49 & 12.41 & 10.62 & 29.81 & 54.78 & 21.58 & 11.60 & 8.06 & 24.01 \\
 & SK2Decompile-6.7B* & 86.59 & 70.59 & 61.31 & 57.52 & 69.00 & 69.76 & 62.33 & 54.83 & 51.58 & 59.63 \\
 & AutoDecompiler-Pscode-1.3B & 95.39 & 93.33 & 92.00 & 85.90 & 91.66 & 69.50 & 59.18 & 52.09 & 46.18 & 56.74 \\
 & AutoDecompiler-Pscode-6.7B & 96.95 & 93.52 & 95.05 & 91.00 & 94.13 & \textbf{74.95} & 65.66 & 60.32 & 58.14 & 64.77 \\
 & AutoDecompiler-Pscode-30B & \textbf{97.71} & \textbf{98.11} & \textbf{100.00} & \textbf{98.85} & \textbf{98.67 }& 73.38 & \textbf{70.25} & \textbf{66.40} & \textbf{63.04} & \textbf{68.27} \\
\midrule
\multirow{6}{*}{R2I} & DeepSeek-V4 & 53.35 & 53.12 & 49.34 & 52.52 & 52.08 & 60.97 & 53.87 & 54.10 & 54.98 & 55.98 \\
 & DeepSeek Harness (DeepSeek-V4) & 58.69 & 63.79 & 62.91 & 66.49 & 62.97 & 55.20 & 61.43 & 62.23 & 63.05 & 60.48 \\
 & DecLLM (DeepSeek-V4)$^{\dagger}$ & 48.81 & 38.78 & 40.57 & 38.47 & 41.66 & 54.55 & 40.46 & 36.42 & 34.63 & 41.51 \\
 & AutoDecompiler-Pscode-1.3B & \textbf{64.43} & 73.12 & \textbf{74.53} & \textbf{73.60} & \textbf{71.42} & \textbf{69.86} & \textbf{75.68} & \textbf{74.07} & \textbf{71.01} & \textbf{72.65} \\
 & AutoDecompiler-Pscode-6.7B & 60.80 & 73.30 & 72.13 & 71.59 & 69.45 & 68.28 & 72.57 & 71.33 & 68.31 & 70.12 \\
 & AutoDecompiler-Pscode-30B & 63.43 & \textbf{73.36} & 72.51 & 71.18 & 70.12 & 58.74 & 67.70 & 65.01 & 62.44 & 63.47 \\
\bottomrule
\end{tabular}%
}
\vspace{-3ex}
\label{ref67result}
\end{table*}

Table~\ref{e2e67result} reports the comparison of traditional decompilers and end-to-end decompilation methods. 
Table~\ref{ref67result} compares pseudo-code-based decompilation methods.
In particular, we assigned GPT-5.6 to end-to-end decompilation and DeepSeek-V4 to pseudo-code-based decompilation.

Next, we present several observations and analyses based on the experimental results.

\noindent\textbullet~
Across both input settings, AutoDecompiler shows strong competitiveness in Re-com and Re-exe against traditional decompilers, specialized decompilation LLMs, advanced general-purpose LLMs, and coding agents.
At the 6.7B scale, AutoDecompiler-E2E achieves an average Re-com of 90.98\% and Re-exe of 72.86\%, outperforming prior best CIM-6.7B by 3.14\% and 14.65\%, respectively; its Re-exe also exceeds GPT-5.6 (57.92\%) and Codex (59.10\%), while traditional decompilers achieve substantially lower results (14.88\% for IDA Pro and 11.89\% for Ghidra).
Similarly, AutoDecompiler-Pscode-6.7B achieves 79.45\% average Re-exe, outperforming SK2Decompile-6.7B (64.32\%), DeepSeek-V4 (66.34\%), and DeepSeek Harness (77.31\%).
These results demonstrate the effectiveness of AutoDecompiler across both end-to-end and pseudo-code-based decompilation settings.
Although the average R2I (68.91\%) of AutoDecompiler-E2E-6.7B is 1.56\% lower than that of CIM-6.7B, it still outperforms the other LLM-based baselines.


\noindent\textbullet~
Model scaling consistently improves behavioral recovery while showing limited benefits for readability.
As AutoDecompiler scales from 1.3B to 6.7B and 30B, its average Re-exe across HumanEval and MBPP increases from 63.56\% to 72.86\% and 78.78\% in the end-to-end setting, and from 74.20\% to 79.45\% and 83.47\% in the pseudo-code-based setting.
In contrast, the corresponding R2I remains around 68.72\%--69.31\% for end-to-end decompilation and remains around 66.80\%--72.04\% for pseudo-code-based decompilation.
Notably, the 1.3B AutoDecompiler already outperforms the 6.7B specialized baselines in average Re-exe in both settings (63.56\% vs.\ 58.21\% for CIM and 74.20\% vs.\ 64.32\% for SK2Decompile), demonstrating strong parameter efficiency.

\noindent\textbullet~
The gains of AutoDecompiler increasingly translate from recompilability into behavioral consistency as the model scales.
The average Re-com--Re-exe gap decreases from 21.71\% to 18.12\% and 12.98\% for the 1.3B, 6.7B, and 30B end-to-end models, respectively, and from 17.14\% to 12.28\% and 8.18\% for their pseudo-code-based counterparts.
This consistent trend indicates that larger AutoDecompiler models more effectively convert recompilable outputs into behaviorally consistent code, aligning with its execution-feedback-driven training objective.

\noindent\textbullet~
Traditional decompilers and LLM-based refinement exhibit strong complementarity.
Averaged across HumanEval and MBPP, IDA Pro alone achieves only 18.10\% Re-com and 14.88\% Re-exe, whereas AutoDecompiler-Pscode-6.7B, which takes decompiler-generated pseudo-code as input, reaches 91.73\% and 79.45\%, respectively.
These results demonstrate the potential of combining traditional decompilation analysis with LLM-based refinement for recovering recompilable and behaviorally consistent code.

\noindent\textbullet~
Despite not being specifically trained for decompilation, general-purpose LLMs demonstrate substantial decompilation capability and achieve results competitive with several specialized models.
Their continued advances may further strengthen LLM-based decompilation, but large-scale deployment still involves concerns about inference cost, efficiency, and data privacy.
This highlights the complementary value of lightweight, domain-specialized models such as AutoDecompiler, which can offer local deployment and also serve as backbone models for agentic systems.

\noindent\textbullet~
General-purpose coding agents can substantially improve their backbone LLMs in some cases, but the gains are highly dataset dependent.
In end-to-end decompilation, Codex improves GPT-5.6 Re-exe on HumanEval from 60.06\% to 66.34\% , but decreases it on MBPP from 55.78\% to 51.86\%.
Similarly, in pseudo-code-based decompilation, DeepSeek Harness improves DeepSeek-V4 from 71.95\% to 93.75\% on HumanEval, but only from 60.73\% to 60.87\% on MBPP.
This suggests that generic agentic refinement can be effective, but does not provide consistent gains across datasets.

\noindent
\fbox{%
\parbox{\dimexpr\linewidth-2\fboxsep-2\fboxrule\relax}{%
\textbf{Summary:}
Overall, AutoDecompiler shows strong competitiveness across all baseline categories.
It substantially outperforms traditional decompilers and SOTA decompilation LLMs in recompilability and behavioral recovery, while remaining competitive with advanced general-purpose LLMs and coding agents.
}%
}



\begin{table*}[h!]
\caption{Re-exe performance comparison across training stages under the same three-turn refinement. (\%)}
\centering
    \resizebox{\textwidth}{!}{
   \scriptsize
\begin{tabular}{c|ccccccccccc}
\toprule

\multirow{2}{*}{Size} & \multirow{2}{*}{Method} & \multicolumn{5}{c}{HumanEval} & \multicolumn{5}{c}{MBPP} \\
\cmidrule(lr){3-7} \cmidrule(lr){8-12}
 &  & O0 & O1 & O2 & O3 & AVG & O0 & O1 & O2 & O3 & AVG \\
\midrule
\multirow{3}{*}{1.3B} & Vanilla & 3.92 & 6.67 & 2.67 & 2.96 & 4.06 & 6.19 & 7.10 & 9.11 & 5.31 & 6.93 \\
 & SFT & 83.67 & 62.59 & 56.08 & 53.10 & 63.86 & 67.25 & 55.15 & 47.26 & 43.50 & 53.29 \\
 & AutoDecompiler &\textbf{ 95.39} & \textbf{93.33} & \textbf{92.00 }& \textbf{85.90} & \textbf{91.66} & \textbf{69.50 }& \textbf{59.18} & \textbf{52.09} & \textbf{46.18} & \textbf{56.74} \\
\midrule
\multirow{3}{*}{6.7B} & Vanilla & 67.59 & 60.87 & 49.30 & 42.11 & 54.97 & 60.11 & 56.70 & 47.03 & 34.86 & 49.68 \\
 & SFT & 88.16 & 71.03 & 67.36 & 64.58 & 72.78 & 74.43 & 62.57 & 56.29 & 55.63 & 62.23 \\
  & AutoDecompiler & \textbf{96.95} & \textbf{93.52} & \textbf{95.05} & \textbf{91.00} & \textbf{94.13} & \textbf{74.95} & \textbf{65.66} & \textbf{60.32} & \textbf{58.14} & \textbf{64.77} \\
\midrule
\multirow{3}{*}{30B} & Vanilla & 76.62 & 65.36 & 66.23 & 56.77 & 66.25 & 66.21 & 59.57 & 49.52 & 41.12 & 54.11 \\
 & SFT & 85.62 & 76.16 & 72.67 & 73.83 & 77.07 & 70.28 & 65.95 & 59.13 & 56.93 & 63.07 \\
 & AutoDecompiler & \textbf{97.71} & \textbf{98.11} & \textbf{100.00} & \textbf{98.85} & \textbf{98.67} & \textbf{73.38} & \textbf{70.25} & \textbf{66.40} & \textbf{63.04} & \textbf{68.27} \\
 
\bottomrule
\end{tabular}%
}
\vspace{-3ex}
\label{MTRLEffectiveness}
\end{table*}

\subsection{Effectiveness of Multi-turn Reinforcement Learning (RQ2)}

As described in Section~\ref{sec:framework}, AutoDecompiler is constructed following the common paradigm of domain-specialized LLMs, where a general code LLM is progressively adapted to the decompilation task through task-oriented specialization. 
To answer RQ2, we compare three training stages of AutoDecompiler under the pseudo-code-based setting: the vanilla base model, the SFT model, and the final model trained with multi-turn RL.
Importantly, all three variants are evaluated using the same three-turn refinement setting, 
ensuring that their performance differences reflect the effect of training rather than different inference budgets.
We conduct the comparison across three model scales (1.3B, 6.7B, and 30B) on both HumanEval and MBPP, and report Re-exe under four optimization levels (as shown in Table~\ref{MTRLEffectiveness}).

We observe consistent performance improvements across the three training stages under the same three-turn refinement setting.
SFT substantially improves the vanilla model at all model scales, demonstrating the importance of decompilation-specific supervised adaptation.
For example, on HumanEval, the average Re-exe increases from 4.06\% to 63.86\%, from 54.97\% to 72.78\%, and from 66.25\% to 77.07\% for the 1.3B, 6.7B, and 30B models, respectively.

More importantly, multi-turn RL consistently improves upon SFT while using the same three-turn inference budget.
On HumanEval, AutoDecompiler improves the average Re-exe over SFT by 27.80\%, 21.35\%, and 21.60\% for the 1.3B, 6.7B, and 30B models, respectively.
On MBPP, the corresponding gains are 3.45\%, 2.54\%, and 5.20\%.
Notably, AutoDecompiler outperforms its SFT counterpart in all 24 combinations of model scale, dataset, and optimization level.
These results demonstrate that multi-turn RL enables the model to make more effective use of iterative refinement, rather than merely benefiting from additional inference turns.

\noindent
\fbox{%
\parbox{\dimexpr\linewidth-2\fboxsep-2\fboxrule\relax}{%
\textbf{Summary:}
Under the same three-turn refinement setting, multi-turn RL consistently improves performance over SFT across model scales, datasets, and optimization levels, demonstrating that RL training enables more effective use of iterative refinement.
}%
}

\begin{table}[H]
\centering
\caption{Re-exe performance of AutoDecompiler-E2E-1.3B with different ablation variants on the HumanEval dataset. \textbf{com/exe} denotes compilation and execution.
(\%)}
\resizebox{\columnwidth}{!}{
{\fontsize{9pt}{10pt}\selectfont
\begin{tabular}{c|ccccc}
\hline
\textbf{Model} & \textbf{O0} & \textbf{O1} & \textbf{O2} & \textbf{O3} & \textbf{AVG} \\
\cmidrule(lr){1-1}\cmidrule(lr){2-6}
			
Full Model & \textbf{86.24} & \textbf{69.39} & \textbf{65.35} & \textbf{71.76} & \textbf{73.19} \\\hline

w/o C-like validity & 80.83 & 60.94 & 51.59 & 55.00 & 62.09 \\
w/o com/exe correctness& 82.35 & 58.40 & 52.07 & 54.39 & 61.80 \\
w/o syntactic reward & 82.50 & 62.60 & 55.17 & 60.19 & 65.12 \\
w/o semantic reward& 79.51 & 55.47 & 52.00 & 56.14 & 60.78 \\

w/o trajectory reward & 80.67 & 56.59 & 53.28 & 53.57 & 61.03 \\

\hline
\end{tabular}
}
}
\label{ablationtable}
\vspace{-2ex}
\end{table}

\subsection{Contribution of Key Components (RQ3)}

To answer RQ3, we conduct ablation studies on AutoDecompiler-E2E-1.3B to evaluate the contribution of each key component.
Different from RQ2, which evaluates multi-turn RL under the pseudo-code-based setting, we perform ablation in the more challenging end-to-end setting, where the model directly takes assembly code as input without relying on pseudo-code generated by traditional decompilers.
We remove each reward component individually and report Re-exe on HumanEval, as shown in Table~\ref{ablationtable}.

Overall, removing any component consistently degrades Re-exe across all optimization levels, confirming that each component contributes to AutoDecompiler.
Among the validation rewards, removing the semantic reward causes the largest performance drop, reducing the average Re-exe from 73.19\% to 60.78\%, a decrease of 12.41\%.
Removing the compilation/execution correctness, C-like validity, and syntactic rewards decreases the average Re-exe to 61.80\%, 62.09\%, and 65.12\%, respectively.
These results show that behavioral correctness, code validity, syntactic structure, and data-flow semantics provide complementary supervision for decompilation.

The progress-aware trajectory reward is also critical to multi-turn refinement.
The \textit{w/o trajectory reward} variant retains only the final-turn reward, removing the intermediate-history and regression-aware terms.
Its average Re-exe drops from 73.19\% to 61.03\%, a decrease of 12.16\%.
This result shows that optimizing only the final outcome is insufficient, while explicitly modeling trajectory quality and penalizing regressions provides important guidance for effective multi-turn refinement.

\noindent
\fbox{%
\parbox{\dimexpr\linewidth-2\fboxsep-2\fboxrule\relax}{%
\textbf{Summary:}
All key components contribute to AutoDecompiler, with semantic supervision and progress-aware trajectory rewarding providing particularly strong gains for behavioral recovery and multi-turn refinement.
}%
}

\begin{table*}[h!]
\caption{Cross-compiler and cross-decompiler generalization performance of pseudo-code-based decompilation methods on HumanEval dataset, measured by Re-exe. (\%)}
\vspace{-1.5ex}
\centering
   \scriptsize

\setlength{\tabcolsep}{2pt}
\renewcommand{\arraystretch}{1.12}
\hspace*{0.015\textwidth}
\begin{minipage}[t]{0.47\textwidth}
\centering
{(a) Cross-compiler: Clang + IDA Pro}\par\smallskip
\begin{tabular*}{\linewidth}{@{\extracolsep{\fill}}cccccc@{}}
\toprule
Model / Method & O0 & O1 & O2 & O3 & AVG \\
\midrule
IDA Pro & 21.34 & 17.07 & 12.80 & 11.59 & 15.70 \\
DeepSeek-V4 & 73.17 & 48.78 & 41.46 & 48.17 & 52.90 \\
AutoDecompiler-Pscode-1.3B & 75.78 & 49.69 & 41.56 & 41.56 & 52.15 \\
AutoDecompiler-Pscode-6.7B & \textbf{85.06} & 63.40 & 56.76 & 55.78 & 65.25 \\
AutoDecompiler-Pscode-30B & 83.23 & \textbf{70.97} &\textbf{ 73.47} & \textbf{73.29} & \textbf{75.24} \\
\bottomrule
\end{tabular*}
\end{minipage}%
\hfill
\begin{minipage}[t]{0.47\textwidth}
\centering
{(b) Cross-decompiler: GCC + Ghidra}\par\smallskip
\begin{tabular*}{\linewidth}{@{\extracolsep{\fill}}cccccc@{}}
\toprule
Model / Method & O0 & O1 & O2 & O3 & AVG \\
\midrule
Ghidra & 19.51 & 14.63 & 12.20 & 11.59 & 14.48 \\
DeepSeek-V4 & 73.78 & 47.56 & 50.00 & 44.51 & 53.96 \\
AutoDecompiler-Pscode-1.3B & 50.62 & 30.19 & 35.19 & 28.21 & 36.05 \\
AutoDecompiler-Pscode-6.7B & 70.78 & 41.45 & 47.40 & 42.67 & 50.58 \\
AutoDecompiler-Pscode-30B & \textbf{77.12} &\textbf{ 58.28} & \textbf{58.94} & \textbf{62.07} &\textbf{ 64.10} \\
\bottomrule
\end{tabular*}
\end{minipage}%
\hspace*{0.015\textwidth}
\vspace{-2ex}
\label{crossgeneralizeresult}
\end{table*}

\subsection{Generalization across Compilation and Decompilation Toolchains (RQ4)}

Our main experiments use GCC for compilation and IDA Pro for pseudo-code generation.
To answer RQ4, we evaluate two out-of-distribution settings on HumanEval: \textit{cross-compiler}, replacing GCC with Clang while retaining IDA Pro, and \textit{cross-decompiler}, replacing IDA Pro with Ghidra while retaining GCC.
Table~\ref{crossgeneralizeresult} reports the corresponding Re-exe results.

AutoDecompiler retains decompilation capability under both unseen toolchains, although performance decreases from the original GCC+IDA Pro setting.
For the 1.3B, 6.7B, and 30B models, average Re-exe drops from 91.66\%, 94.13\%, and 98.67\% to 52.15\%, 65.25\%, and 75.24\% under Clang+IDA Pro, and to 36.05\%, 50.58\%, and 64.10\% under GCC+Ghidra, respectively.
This shows that AutoDecompiler generalizes beyond its training toolchain, while toolchain shifts remain challenging.

Generalization improves with model scale.
The cross-compiler performance drop decreases from 39.51\% to 28.88\% and 23.43\% from 1.3B to 6.7B and 30B, while the cross-decompiler drop decreases from 55.61\% to 43.55\% and 34.57\%.
The 30B achieves the best average Re-exe in both settings, reaching 75.24\% and 64.10\%, respectively, and outperforming DeepSeek-V4 by 22.34\% and 10.14\%.

Cross-decompiler generalization is consistently more challenging than cross-compiler generalization.
For the three model sizes, GCC+Ghidra yields average Re-exe scores 16.10\%, 14.67\%, and 11.14\% lower than Clang+IDA Pro.
This suggests that changing the decompiler has a larger impact on pseudo-code-based decompilation than changing the compiler. Meanwhile, the decreasing performance gap between the two settings with increasing model size suggests that larger models are more robust to toolchain variations.

\begin{figure*}
  \centering
  \includegraphics[height=2.4in,width=7.0in]{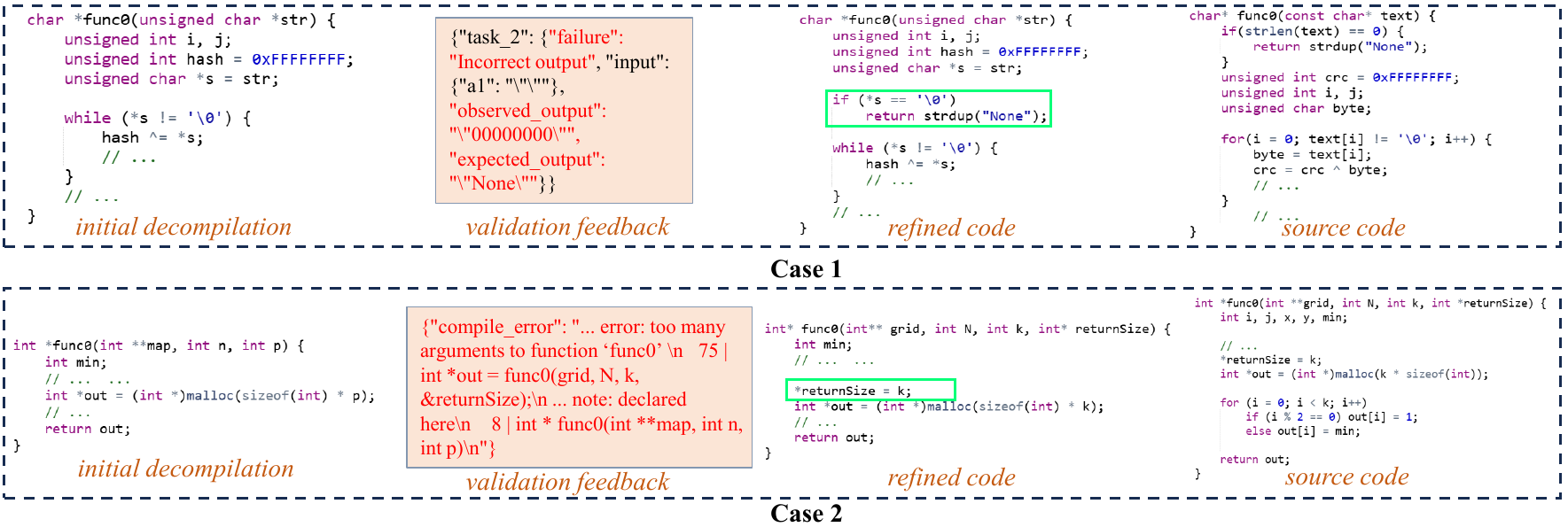}
  \caption{Case studies of AutoDecompiler repairing incorrect decompiled code using validation feedback.}
  \label{casestudyfig}
  \vspace{-3ex}
\end{figure*}

\subsection{Case Study (RQ5)}

To answer RQ5, Figure~\ref{casestudyfig} presents two representative cases showing how AutoDecompiler repairs decompilation errors using validation feedback.
The reference source code is shown only for comparison and is not available to the model during refinement.

\textbf{Case 1: repairing a behavioral error.}
The initial decompilation recovers the main computation but misses the empty-input branch, producing \texttt{"00000000"} instead of the expected \texttt{"None"}.
The execution feedback provides the failing input together with the observed and expected outputs.
AutoDecompiler accordingly introduces an empty-input check and returns \texttt{strdup("None")}, recovering the missing boundary-case behavior.
This case shows that execution feedback can guide targeted repair of input-specific behavioral errors.

\textbf{Case 2: repairing a compilation error.}
The initial decompilation produces an incomplete function signature that omits the \texttt{returnSize} parameter, causing a ``too many arguments'' compilation error.
Guided by the compiler diagnostics, AutoDecompiler restores \texttt{returnSize}, assigns \texttt{*returnSize = k}, and adjusts the output allocation accordingly.
This case demonstrates that compiler feedback can guide the repair of interface errors and their associated program logic.

\noindent
\fbox{%
\parbox{\dimexpr\linewidth-2\fboxsep-2\fboxrule\relax}{%
\textbf{Summary:}
AutoDecompiler exploits different validation signals to perform targeted refinement, using execution feedback to repair behavioral errors and compiler diagnostics to correct interface and compilation errors.
}%
}

\section{Discussion and Limitations}
\label{sec:discussion}

Comparisons with baselines sharing the same DeepSeek-Coder backbone help control for potential benchmark exposure during pre-training, making the observed gains more attributable to AutoDecompiler's task-specific training and multi-turn RL. Moreover, its feedback-driven training framework can potentially extend to other binary-analysis and reverse-engineering tasks.

Although AutoDecompiler achieves performance comparable to other decompilation LLMs with a substantially smaller training dataset, several limitations remain.

First, the number of refinement turns is constrained by the context length of the underlying LLM. For the DeepSeek-Coder-based models, only a limited history of generated code and diagnostic feedback can be retained, which restricts deeper multi-turn repair. Longer-context models or more effective history selection strategies may further improve iterative refinement.

Second, AutoDecompiler is currently designed at the single-function level. While this setting follows common decompilation benchmarks, it does not fully exploit interprocedural information, such as caller-callee relationships, global variables, shared data structures, and cross-function type dependencies. Incorporating such program-level context is an important direction for future work.

Third, although RQ4 demonstrates that AutoDecompiler generalizes beyond the GCC+IDA Pro training toolchain, its performance still degrades under unseen toolchains, particularly when changing the decompiler. This indicates that robustness remains constrained by training-data and toolchain diversity. Expanding coverage to more compilers, decompilers, optimization settings, architectures, software domains, and source-code styles may improve generalization. Its effectiveness on heavily stripped, obfuscated, packed, or protected binaries also remains unexplored.

Finally, LLM-based decompilation inevitably faces the risk of generating hallucinated code, which may be readable or syntactically valid but inconsistent with the original binary semantics. Reducing such hallucinations remains an important research direction for improving the reliability of decompilation LLMs.

\section{Conclusion}

In this paper, we presented AutoDecompiler, an RL-trained LLM that reframes binary decompilation as feedback-driven multi-turn refinement. AutoDecompiler uses compilation, execution, and I/O feedback, together with multi-dimensional rewards, stage-aware diagnostics, and progress-aware trajectory rewarding, to iteratively improve generated code. Experiments across input settings, model scales, and benchmarks demonstrate improved recompilability and re-executability over vanilla and single-turn baselines, while ablation and case studies validate its key designs and practical repair capability.

\newpage

\section*{Data Availability}
The complete training and evaluation code and dataset, along with the Appendix of this paper, are archived on Zenodo at
\url{https://doi.org/10.5281/zenodo.23120286}.

We open-source the AutoDecompiler family to facilitate community research and applications at HuggingFace: \url{https://huggingface.co/AutoDecompiler}.

\section*{Ethics Considerations}\label{EthicsConsiderations}

This study does not involve any human subjects research.

This study does not involve the disclosure of any software vulnerabilities.

The training and evaluation data are obtained from publicly available and permissively licensed sources, including open-source software projects. We carefully screen the data to avoid sensitive or proprietary content.


\section*{LLM Usage Considerations}

LLMs were used for editorial purposes in this manuscript, and all outputs were inspected by the authors to ensure accuracy and originality.

For all compared approaches, we follow the implementation details and experimental settings reported in the original publications and reproduce them using the designated open-source or proprietary LLMs. This practice helps ensure the fairness and reliability of the experimental comparison.

Our research focuses on developing LLMs for binary decompilation. We adopt publicly available code-oriented foundation LLMs, including DeepSeek-Coder and Qwen3-Coder, as the backbone architectures. The training and evaluation data are collected from publicly accessible and permissively licensed sources, such as open-source software projects. We carefully screen the data to avoid the inclusion of sensitive or proprietary content, and all experiments are conducted using resources intended for research and educational purposes.

\bibliographystyle{IEEEtran} 
\bibliography{refers}

\newpage
\appendix

\section{Method}
\subsection{SFT Dataset}
\label{appendsft}

\begin{tcolorbox}[
title={Pseudo-code-based Decompilation Prompt},
colback=gray!5,
colframe=black!60,
fonttitle=\bfseries
]
\textbf{System message.}

You are a decompilation expert. Your task is to analyze the pseudo-code generated by an existing decompilation tool and generate accurate high-level C source code. The generated C code should have:
(1) correct program logic and control-flow structure;
(2) properly typed and named functions;
(3) properly typed and named variables and data; and
(4) standard and readable C syntax.

\medskip
\textbf{User message.}

Please understand the provided pseudo-code and decompile it into the corresponding high-level C source code.

\medskip
\texttt{- The pseudo-code:}

\begin{verbatim}
<PSEUDO_CODE>
\end{verbatim}

The high-level C source code is:
\end{tcolorbox}

\subsection{RL Dataset}
\label{appendrl}

The core elements of an RL training example are presented at Table~\ref{tab:appendrldata} in this section.

\begin{table}[t]
\centering
\caption{Core elements of an RL training sample.}
\label{tab:appendrldata}
\small
\begin{tabular}{p{0.22\linewidth} p{0.68\linewidth}}
\toprule
\textbf{Key} & \textbf{Example value} \\
\midrule
\textbf{\texttt{sourcecode}} 
& Ground-truth C function, e.g., \texttt{func0(float numbers[], int size, float threshold)}, which checks whether two array elements differ by less than a given threshold. \\

\textbf{\texttt{pscode}} 
& IDA Pro-generated pseudo-code, e.g., \texttt{\_\_int64 \_\_fastcall func0(\_\_int64 a1, int a2, float a3)}, where pointer arithmetic and recovered local variables are used to approximate the original source logic. \\

\textbf{\texttt{assembly}} 
& Disassembled x86\_64 assembly code, e.g., \texttt{push rbp; mov rbp, rsp; movss [rbp+var\_20], ...; retn}. \\

\textbf{\texttt{dependencies}}
& External dependencies required to recompile and execute the generated code, such as header files, library declarations, global variables, and type definitions. For example, this sample requires \texttt{\#include <stdio.h>}, \texttt{\#include <stdlib.h>}, and \texttt{\#include <math.h>}. \\

\textbf{\texttt{testcases}}
& Input-output test cases used to validate the execution behavior of the generated code, e.g., \texttt{assert(func0(b, 5, 0.95) == 1)} and \texttt{assert(func0(b, 5, 0.8) == 0)}. \\

\bottomrule
\end{tabular}
\end{table}

\section{Experiment}
This section provides additional details that complement Section \ref{mainexpes} in the main text.
\subsection{Experimental Setups}




\subsubsection{Baseline LLMs}
\label{appbasellms}
\quad\\

We further introduce the details of the baseline LLMs used in our evaluation.

\begin{itemize}

\item \textbf{GPT-5.6}~\cite{openai2026gpt56} is a representative general-purpose LLM with strong capabilities in code understanding and generation. We include it to evaluate how a powerful closed-source foundation model performs on decompilation without task-specific fine-tuning. We choose the GPT-5.6-luna version as the baseline.

\item \textbf{DeepSeek-V4}~\cite{deepseekai2026deepseekv4} is a large-scale open-source language model with strong capabilities in natural language understanding, code generation, and reasoning tasks. We evaluate the DeepSeek-V4-flash version as a representative general-purpose open-source LLM baseline.

\item \textbf{Codex}~\cite{chen2021codex} is OpenAI's coding agent designed to complete software-engineering tasks through interaction with a development environment. It can inspect and modify code, execute commands and tests, observe execution feedback, and iteratively revise its solutions until task completion. In our evaluation, we use Codex with GPT-5.6 as the underlying LLM.

\item \textbf{DeepSeek Harness}~\cite{deepseekharness2026} is an open-source agent harness developed by DeepSeek for coding and general task completion. It equips the underlying LLM with tools for file editing, shell execution, search, planning, and workflow execution, enabling the agent to iteratively modify code based on intermediate execution and tool feedback. In our evaluation, we use DeepSeek Harness with DeepSeek-V4 as the underlying LLM.


\item \textbf{DecLLM}~\cite{Wong2025DecLLM} is an iterative LLM-based repair framework that leverages compilation errors and runtime feedback to transform decompiler outputs into recompilable code, enabling downstream programmatic analysis such as CodeQL-based vulnerability detection.

\item \textbf{FAE}~\cite{feng2024sc2dec} improves decompilation by leveraging debugging information to construct fine-grained alignment between assembly code and source statements. It is built by further fine-tuning DeepSeek-Coder-based LLM4Decompile-End-6.7B using 10K training samples.

\item \textbf{LLM4Decompile-DCBench}~\cite{tan2025decombench} is a family of end-to-end decompilation models obtained by further fine-tuning the 1.3B and 6.7B DeepSeek-Coder-based LLM4Decompile-End checkpoints on real-world binary--source function pairs from Decompile-Bench~\cite{tan2025decombench}. We use the better-performing 6.7B variant as our baseline.

\item \textbf{CIM}~\cite{Liu2025TheCS} improves the structural and data object recovery capability of decompilation LLMs by incorporating control-flow information and data mapping. It develops two decompilation LLMs with 1.3B and 6.7B parameters, both of which are built upon DeepSeek-Coder-based LLM4Decompile-End and incrementally fine-tuned on 8.96 million samples.


\item \textbf{Idioms}~\cite{Dramko2025IdiomsND} jointly recovers source code and user-defined composite types by leveraging interprocedural context. 

\item \textbf{SK2Decompile}~\cite{Tan2025SK2DecompileLT} is a two-phase LLM-based decompilation framework that first recovers source-level program structure as an identifier-obfuscated intermediate representation and then restores meaningful identifiers, enabling separate optimization of functional correctness and code readability. The LLM is obtained by further fine-tuning DeepSeek-Coder-based  LLM4Decompile-ref on 5 million training samples.


\end{itemize}

\end{document}